\documentstyle[prl,aps,epsf]{revtex}

\begin{document}
\draft
\title{
Anisotropic Transport of Two-Dimensional Holes
in High Landau Levels
}
\author{M. Shayegan, H. C. Manoharan,\cite{HCMauthor}, S. J. Papadakis,
and E.
P. De Poortere}
\address{Department of Electrical Engineering, Princeton University,
    Princeton, New Jersey 08544}

\maketitle
\begin{abstract}
Magnetoresistance data taken along $[\bar{2}33]$ and $[01\bar{1}]$ directions in
a GaAs/AlGaAs two-dimensional hole sample with van der Pauw geometry exhibit
significant anisotropy at half-integer filling factors.  The anisotropy appears
to depend on both the density and symmetry of the hole charge distribution.
\end{abstract}
\vspace{5mm}
\hspace{1.56cm}Keywords:  Quantum Hall Effect, Anisotropy

A remarkable magnetotransport
anisotropy at half-integer fillings was recently reported in high-mobility
GaAs/AlGaAs two-dimensional electron systems (2DESs) \cite{Lilly99,Du99}.
In particular, at half fillings in the third and higher Landau levels ($\nu \ge
\frac92$), the in-plane longitudinal magnetoresistance ($R_{xx}$) in one
direction was observed to be much larger than in
the perpendicular direction.  We present here qualitatively
similar anomalies in a high-mobility GaAs/AlGaAs 2D {\it hole} system (2DHS).
The 2DHS data, however, provide an intriguing twist to this problem as they
exhibit the anomaly at fillings as small as $\nu=\frac52$.

\begin{figure}
\epsfxsize=13cm \epsfbox{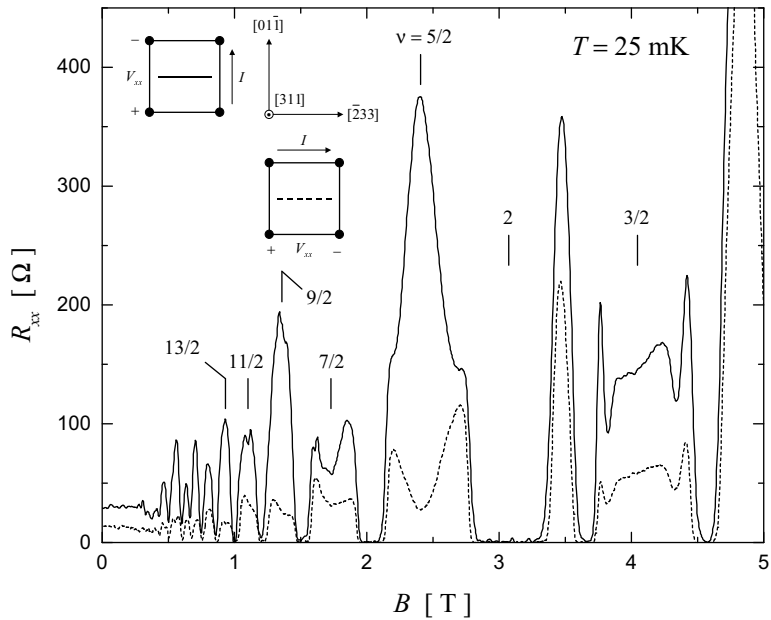} \caption{Longitudinal
magnetoresistance of a 2DHS measured along perpendicular
directions (see insets for van der Pauw measurement geometry). The
sample is cooled to 25 mK and has an areal hole density of $1.5
\times 10^{11}$ cm$^{-2}$.} \label{fig1}
\end{figure}

Figure \ref{fig1} highlights our data which were taken with the magnetic field
$B$ perpendicular to the 2DHS.  The sample is a GaAs quantum well flanked by
AlGaAs barriers,
grown on a GaAs (311)A substrate, and modulation-doped with Si.  The van
der Pauw geometry of the sample and the measurement configurations are
shown as insets to the main figure.  The anisotropy of $R_{xx}$ at half filling
$\nu=\frac52$ is most pronounced:
the resistance along the $[01\bar1]$ direction (solid trace)
exhibits a {\it maximum} which is about
15 times larger than the resistance
{\it minimum} observed along $[\bar233]$ (dashed trace).  Note in Fig.
\ref{fig1} that, as $B$ decreases, the strength of the anisotropy
diminishes in an alternating fashion: larger anisotropies are observed at
$\nu=\frac52$, $\frac92$,  and $\frac{13}{2}$ compared to $\frac72$,
$\frac{11}{2}$, and $\frac{15}{2}$.  This is similar to the case of 2DESs
\cite{Lilly99,Du99}.

In Fig. \ref{fig1}, the transport anisotropy persists down to $B = 0$, implying
a higher mobility (by a factor of about two) along $[\bar233]$ compared to
$[01\bar1]$.  This is typical of 2DHSs grown on GaAs (311)A substrates
\cite{Heremans94}.  The origin of this mobility anisotropy can be traced to
corrugations, aligned along $[\bar233]$, which are present at the GaAs/AlGaAs
(311)A interface \cite{Wassermeier95}.  Such interface morphology anisotropy,
however, cannot by itself explain the much larger {\em and} alternating
transport anisotropy observed at high $B$.

Both Refs. \cite{Lilly99} and \cite{Du99} suggest that the anisotropy observed
at half fillings is intrinsic to very high mobility 2DESs and may signal the
formation of a new, correlated, striped phase of the 2D electrons.  They also
conclude that, in the absence of a parallel $B$, such a state forms exclusively
in the higher Landau levels (LLs), namely $N = 2$ and higher, since they observe
the transport anisotropy only at half-fillings $\nu \ge \frac92$.  If so, why
does the 2DHS exhibit the anisotropy at $\nu$ as small as $\frac52$?  The origin
of this difference is not obvious.  We propose here three possibilities:  (i)
The (311)A interface corrugations may help stabilize the anisotropic state in
2DHSs.  (ii)  It may be related to the nonlinear LL fan diagram for 2D holes
and, in particular to the mixing and crossings of the spin-split LLs
\cite{SeeEkenberg85}.  The exact sequence and spin-character of the 2DHS LLs,
however, depend on the 2D density as well as the shape and symmetry of the
confinement potential and the hole wavefunction, so that a quantitative
assesment of this hypothesis requires further work.  (iii)  The larger effective
mass of GaAs holes renders the 2DHS effectively more dilute \cite{Santos92}.  At
a given filling, such diluteness favors crystalline states of the 2D system over
the fractional quantum Hall (FQH) liquid states \cite{Santos92,revQHE}.  It is
therefore possible that a striped phase is favored in the 2DHS at $\nu =
\frac52$ while in the 2DESs the ground state is a FQH liquid.  Such a scenario
is consistent with the very recent data in 2DESs \cite{Pan99,Lilly99b} which
reveal that applying an in-plane $B$-field destroys the $\nu = \frac52$ FQH
state and stabilizes an anisotropic state.

\begin{figure}
\centerline{
  \epsfxsize=15cm
  \epsfbox{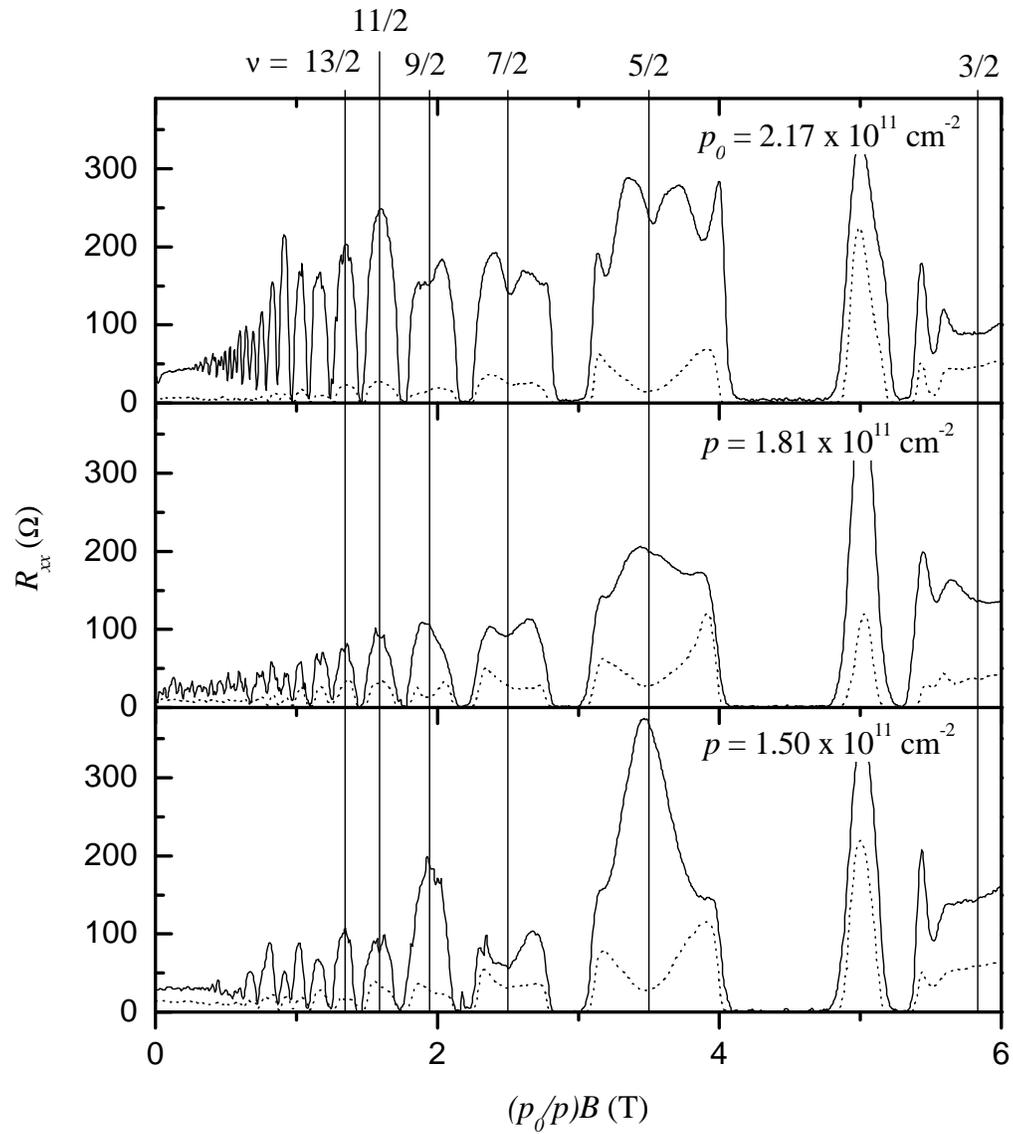}
} \caption{Evolution with denisty of $R_{xx}$ in the 2DHS of Fig.
1.  The data were taken at $T$ = 25 mK.} \label{fig2}
\end{figure}

Figure \ref{fig2} illustrates the evolution with density of $R_{xx}$ for the
2DHS of Fig. \ref{fig1}.  The density obtained by cooling the sample in the
dark, $p_0 = 2.17
\times 10^{11}$ cm$^{-2}$ was lowered via successive illumination at low
temperature by a red light-emitting diode.  At all densities the transport is
highly anisotropic at half-integer fillings.  At the highest density however,
$R_{xx}$ along $[01\bar1]$ exhibits a clear minimum along $\nu = \frac52$.  The
minimum is accompanied by a weak feature in the Hall resistance
\cite{Manoharan94}, suggestive of a developing FQH state.
As the density is lowered, this minimum vanishes and is replaced
by a maximum.  This evolution with decreasing density has some resemblance to
the 2DES data where increasing the in-plane $B$ leads to the destruction of the
$\nu = \frac52$ FQH state \cite{Pan99,Lilly99b}.  It is also consistent with the
above interpretation (iii) that diluteness (lower density) favors the
anisotropic state over the FQH state.

\begin{figure}
\centerline{
  \epsfxsize=15cm
  \epsfbox{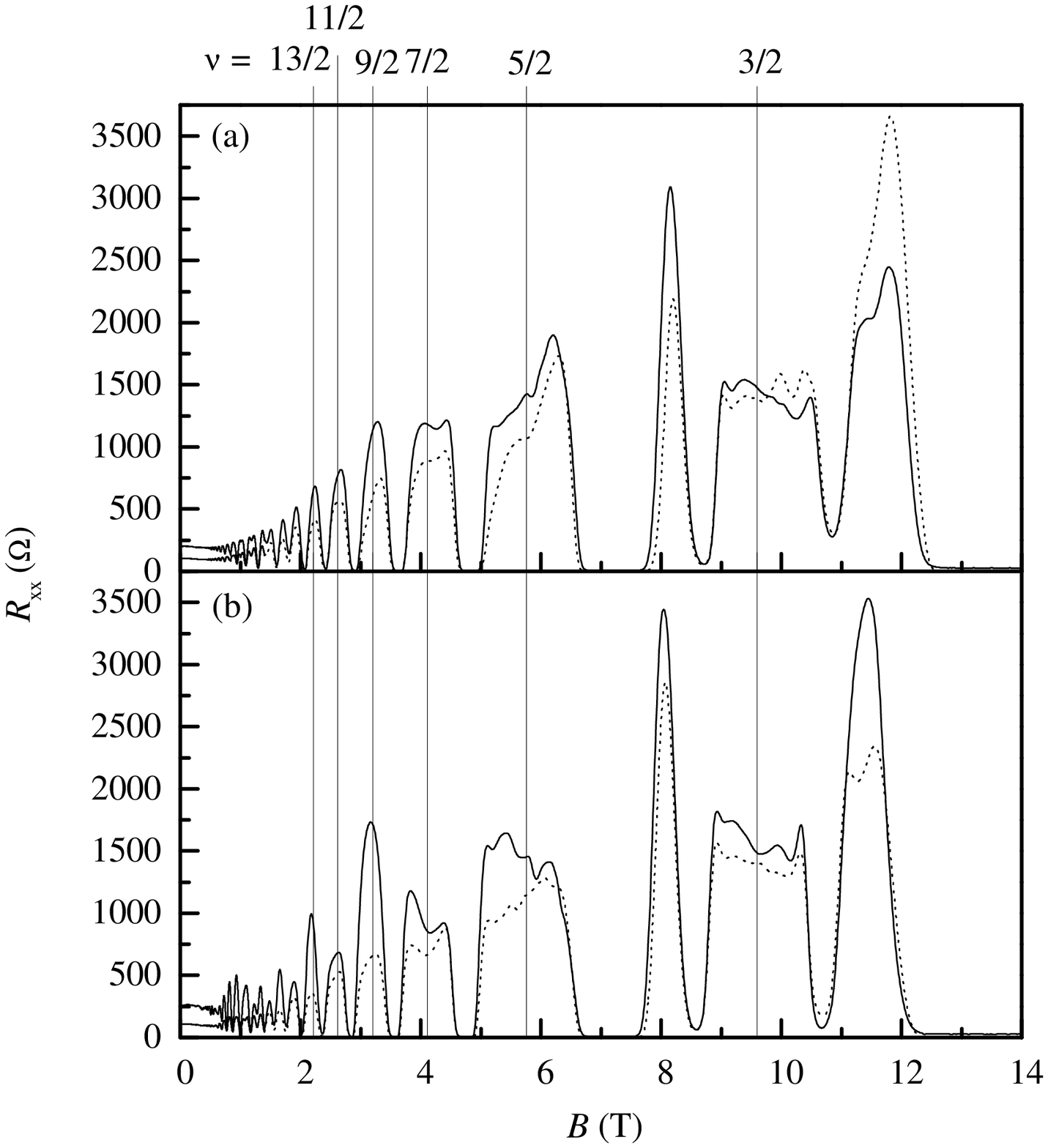}
} \caption{$R_{xx}$ for a 2DHS sample with an L-shaped Hall bar
revealing much less anisotropy than the van der Pauw sample of
Figs. 1 and 2. The density of the sample is fixed at $3.3 \times
10^{11}$ cm$^{-2}$ while the confinement potential for the holes
is varied:  in (a) the potential is symmetric but in (b) it is
made asymmetric by applying a perpendicular electric field of
about 6000 V/cm.  The data were taken at $T$ = 25 mK.}
\label{fig3}
\end{figure}

Finally, in Fig. \ref{fig3} we show data taken on an L-shaped Hall bar sample
from a
different wafer \cite{otherdata}.  The two arms of the Hall bar are aligned
along the $[\bar233]$ and $[01\bar1]$ directions, and the sample has front and
back gates which are used to tune the symmetry
of the hole wavefunction while keeping the density fixed \cite{Papadakis99}.
The data overall exhibit much less transport anisotropy, confirming that the van
der Pauw geometry exagerates anisotropy because of the non-uniform current
distribution \cite{Simon99}.  Moreover, only the data of Fig \ref{fig3}b, taken
with an asymmetric charge distribution, show significant anisotropy at
half-integer fillings for $\nu \ge \frac92$; note also that the magnitude of the
anisotropy alternates with decreasing $B$. The data for the symmetric charge
distribution
(Fig. \ref{fig3}a), on the other hand, exhibit a smaller size anisotropy which
appears to be monotonic with $B$.  While we do not understand the origin of
these differences, we mention that the LL fan diagrams for the two charge
distributions are likely to be different and also the 2DHS becomes more
sensitive to interface corrugations when the charge distribution is made
asymmetric.

This work was supported by the National Science Foundation.

\begin{thebibliography}{10}

\bibitem[\dagger]{HCMauthor}
Current address: IBM Research Division, Almaden Research Center,
650 Harry
  Road, San Jose, CA 95120.

\bibitem{Lilly99}
M.~P. Lilly {\it et~al.}, \prl {\bf 82},  394  (1999).

\bibitem{Du99}
R.~R. Du {\it et~al.}, Solid State Commun. {\bf 109},  389
(1999).

\bibitem{Heremans94}
J.~J. Heremans, M.~B. Santos, K. Hirakawa, and M. Shayegan, J.
Appl. Phys. {\bf
  76},  1980  (1994).

\bibitem{Wassermeier95}
M. Wassermeier {\it et~al.}, \prb {\bf 51},  14721  (1995).

\bibitem{SeeEkenberg85}
See, {\it e.g.}, U. Ekenberg and M. Altarelli, Phys. Rev. B {\bf
32}, 3712
  (1985).

\bibitem{Santos92}
M.~B. Santos {\it et~al.}, \prl {\bf 68},  1188  (1992).

\bibitem{revQHE}
For a review, see M. Shayegan in "Perspectives in QHEs", edited by
A. Pinczuk
  and S. Das Sarma (Wiley, New York, 1997, p. ).

\bibitem{Pan99}
W. Pan {\it et al.}, cond-mat/9903160.

\bibitem{Lilly99b}
M. P. Lilly {\it et al.}, cond-mat/9903196.

\bibitem{Manoharan94}
H.~C. Manoharan and M. Shayegan, \prb {\bf 50},  17662  (1994).

\bibitem{otherdata}
Data taken at $p = 2.3 \times 10^{11}$ cm$^{-2}$ in a Hall bar
sample
  fabricated from the same wafer as in the sample of Figs. 1 and 2 also show
  very little anisotropy.

\bibitem{Papadakis99}
S.~J. Papadakis {\it et~al.}, Science {\bf 283},  2056  (1999).

\bibitem{Simon99}
S. H. Simon, cond-mat/9903086.

\end{thebibliography}
\end{document}